# Mapping the Sun's coronal magnetic field using the Zeeman effect

Thomas A. Schad,[1*] Gordon J.D. Petrie,[2] Jeffrey R. Kuhn,[3] Andre Fehlmann,[1] Thomas Rimmele,[2] Alexandra Tritschler,[2] Friedrich Woeger,[2] Isabelle Scholl,[1] Rebecca Williams,[4] David Harrington,[1] Alin R. Paraschiv,[2] Judit Szente[5]

[1]*National Solar Observatory, 22 'Ōhi'a Kū Street, Makawao, HI 96768, USA*, [2]*National Solar Observatory, 3665 Discovery Drive, Boulder, CO, 80303, USA.*, [3]*University of Hawai'i at Manoa Institute for Astronomy, 34 'Ōhi'a Street, Makawao, HI 96768, USA*, [4]*Observatory Sciences Ltd, 1 New Road, St Ives, PE27 5BG, UK*, [5]*Department of Climate and Space Sciences and Engineering, University of Michigan, Ann Arbor, MI 48109, USA.*

∗*To whom correspondence should be addressed; email: tschad@nso.edu*

## Abstract

Regular remote sensing of the magnetic field embedded within the million-degree solar corona is severely lacking. This reality impedes fundamental investigations of the nature of coronal heating, the generation of solar and stellar winds, and the impulsive release of energy into the solar system via flares and other eruptive phenomena. Resulting from advancements in large aperture solar coronagraphy, we report unprecedented maps of polarized spectra emitted at 1074 nm by $Fe^{+12}$ atoms in the active corona. We detect clear signatures of the Zeeman effect that are produced by the coronal magnetic field along the optically thin path length of its formation. Our comparisons with global magnetohydrodynamic models highlight the valuable constraints that these measurements provide for coronal modeling efforts, which are anticipated to yield subsequent benefits for space weather research and forecasting.

## Teaser

Advanced solar telescope coronagraphically resolves polarization signals generated by the Sun's coronal magnetic field.

## MAIN TEXT

### Introduction

The predominance of the Sun's magnetic field in shaping its outer atmosphere is fantastically evident during a total solar eclipse. On April 8th of this year, eclipse onlookers across North America caught a glimpse of the solar corona near the maximum phase of the solar cycle, when coronal streamers trace the magnetic field's extension into the inner solar system. Here, million-degree temperatures anomalously exist, driven by the transport of magnetoconvective energy originating from the solar interior (1). It is a tenuous region having mass densities eight orders of magnitude lower than the visible solar surface. As a result, radiative diagnostics of the corona are concentrated at ultraviolet, X-ray, and radio wavelengths (2, 3). Only during brief eclipses, or through the use of coronagraphic methods, is the bright solar disk sufficiently blocked to allow high contrast visible and infrared observations of the Sun's corona (4, 5).

Owing to the less rapid decline in the magnetic field amplitude versus height than the plasma density, the Maxwell stresses in the lower corona are generally not substantially opposed by plasma pressure or inertial forces but are mostly contained within the field. Magnetic flux, Poynting flux, and magnetic helicity continuously emerge from the interior into the corona, where Alfvén's frozen flux theorem applies to an excellent approximation in this highly conducting plasma (6). The free energy and stresses cannot be substantially relieved by dissipation except at current sheets (7) and are instead trapped within the magnetic field, building up gradually until they finally escape spectacularly through flares and/or coronal mass ejections (8, 9). The field strength and configuration are paramount in driving these eruptions; however, observations of their state remain largely unavailable. Whereas in the lower atmosphere, polarimetric measurements of the Zeeman effect (10) on visible and infrared spectral lines have facilitated mapping of the photospheric magnetic field for decades (11–13), applying these methods in the upper solar atmosphere is considerably more challenging; although, a number of key advancements have been made in recent years (14) including within the near ultraviolet (15, 16).

In the absence of regular coronal magnetic field measurements, coronal and heliospheric research relies heavily on the photospheric boundary magnetic field to model the outer solar atmosphere (17). This approach has many merits and forms the basis for operational space weather forecasting (18). Yet, these methods are undermined by a deficient knowledge of the extremely complex Maxwell stresses in and throughout the photosphere and the complex partially-ionized chromosphere, which modify the magnetic field before it reaches the corona (19–21). Data-driven simulations may help make this problem more tractable (22–25); though, the lack of coronal magnetic field measurements also inhibits the validation of such methods. Adding to these challenges, radiative emission in the corona is mostly optically thin, meaning a large path length through the highly structured coronal plasma may contribute to observable signals.

Of the potential techniques for coronal magnetic field measurements, polarimetric observations of the Zeeman effect on emission lines offer a comparatively simple diagnostic (26). Yet, such measurements are often thought inaccessible as the great coronal temperature Doppler broadens the emission lines thereby diluting the amplitude of the emitted Zeeman signature. Due to the weak coronal magnetic field amplitude, the targeted signal is only a few to tens of parts per billion of the solar disk's spectral radiance. A few past efforts have demonstrated the possibility of Zeeman measurements using circular polarized observations but have lacked sufficient spatial and temporal resolution to be routinely used in practice (27–29). This landscape is quickly changing due to the advancements afforded by the US National Science Foundation's Daniel K. Inouye Solar Telescope (DKIST, 30). DKIST is a 4-meter aperture off-axis Gregorian solar telescope (the world's largest) inaugurated in Hawaiʻi in 2022 that provides capabilities for large aperture coronagraphic polarimetry. Here, we report the first spatial maps acquired by DKIST of the magnetic field induced Zeeman effect in the off-limb corona. These new capabilities promise to open a new field of routine Zeeman-based coronal magnetometry.

**Results**

*Zeeman-Induced Circular Polarization*

On 22 June 2023 DKIST acquired spectropolarimetric observations of the $Fe^{+12}$ 1074 nm forbidden emission line using the Cryogenic Near-Infrared Spectropolarimeter (CryoNIRSP, 31). The $Fe^{+12}$ line is emitted by plasma at temperatures of $10^{6.23\pm0.16}$ K and is one of multiple potential infrared diagnostics of coronal magnetism (32). In the infrared, the detection potential for circular polarization induced by the Zeeman effect is enhanced because the Zeeman line splitting increases as the square of the wavelength whereas the Doppler broadened thermal line width increases only linearly in wavelength. CryoNIRSP uses a rotating crystal birefringent optic and wire grid polarizers to reconstruct the full state of polarization, as encoded in the Stokes parameters (I, Q, U, V), of its long-slit spectrum centered on the $Fe^{+12}$ line. Spatial maps are created by raster scanning the solar image across the slit (please refer to Materials and Methods section). Meanwhile, field stops occult the solar disk at the telescope's primary and secondary foci, and a Lyot pupil stop is deployed just downstream of the secondary mirror to suppress diffracted light from the primary mirror.

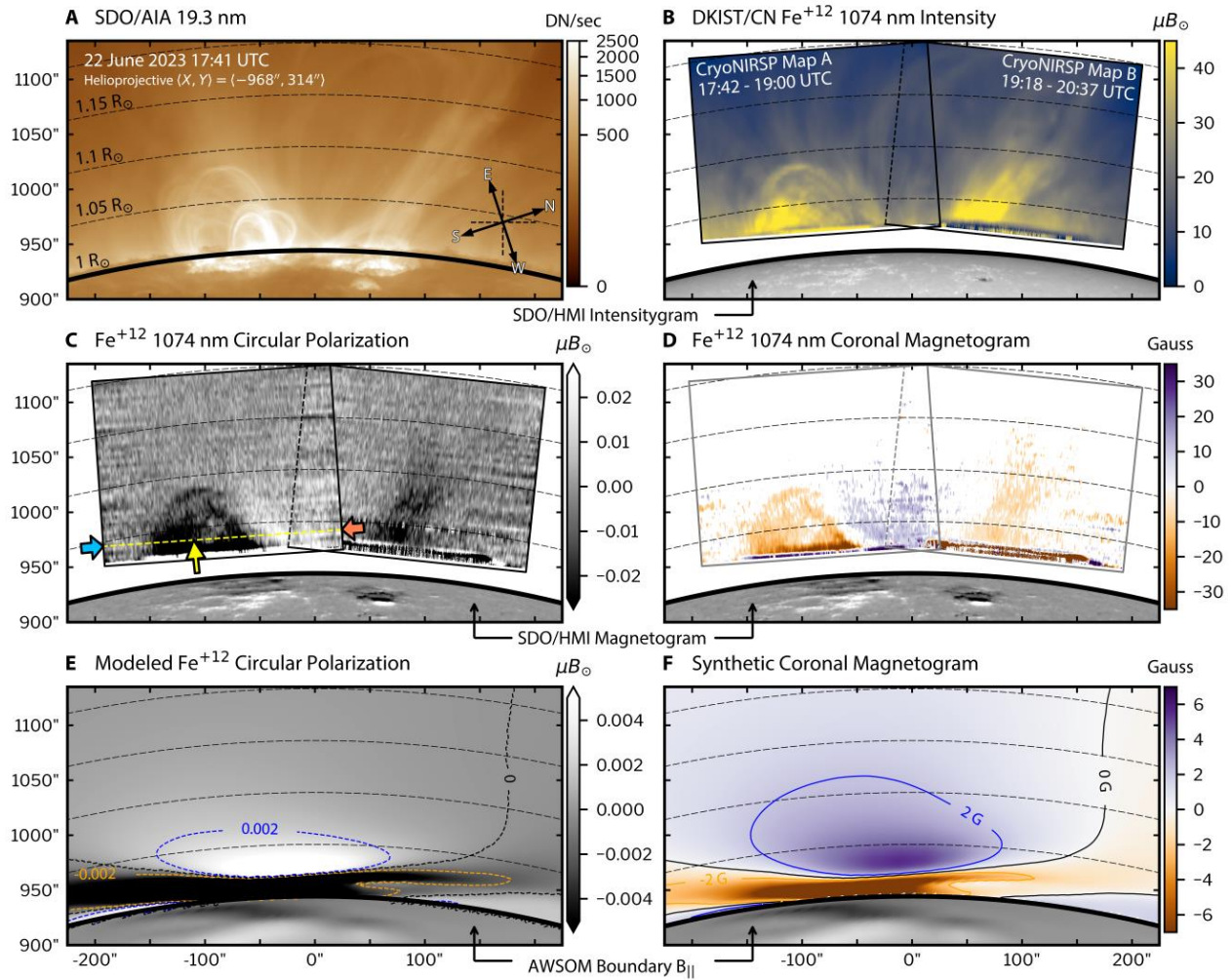

**Fig. 1. The first DKIST coronal magnetogram mapping the magnetic field intensity.** (A) SDO/AIA 19.3 nm image cropped and rotated to geometry of DKIST/CryoNIRSP observations. Vertical coordinates give arcseconds from the center of the solar disk. (B) The peak line amplitude of $Fe^{+12}$ 1074 nm observed within the overlapping raster maps of CryoNIRSP in units of parts per million of the solar disk intensity, *i.e.* $\mu B_\odot$. (C) Peak red-wing amplitude of the measured anti-symmetric circular polarized $Fe^{+12}$ profile. (D) Inferred coronal longitudinal magnetogram in units of Gauss as inferred from the weak field approximation fitted to the circular polarized profiles. (E) Synthetic $Fe^{+12}$ 1074 nm circularly polarized amplitude calculated from the AWSoM coronal model. (F) Synthetic coronal magnetogram inferred from the signal in panel E in the same manner as panel D.

Fig. 1 presents the first maps of circularly polarized coronal emission measured by DKIST and CryoNIRSP. As context, Fig. 1A shows an image of the corona acquired by the Solar Dynamics Observatory's (SDO) Atmospheric Imaging Assembly (AIA, 33, 34) in an ultraviolet bandpass centered at 19.3 nm. $Fe^{+10}$ and $Fe^{+11}$ emission typically dominates this band for quiescent coronal material (35). The active region targeted by DKIST was located on the solar northeast limb, having just rotated into view from the Earth's perspective. While the region exhibited eruptive behavior around 3:45 UT and 15:45 UT, during the CyroNIRSP observations from 17:40 to 20:40 UT, the region's apparent structure was comparatively stable. CryoNIRSP conducted two raster scans with deep exposures, labeled A and B in Fig. 1B (36, 37), during which its 0.5 arcsec wide slit was oriented parallel to the solar limb and sequentially scanned outwards to cover a 180 x 225 arcsec field of view in each map. At each of forty-six raster positions per map, the cumulative integration time was 81.2 seconds. Over the 88-minute duration of each map, the spectrograph acquired 51,520 camera frames, which are processed and calibrated according to the methods described below.

These CryoNIRSP observations successfully resolve the spatial distribution of the $Fe^{+12}$ 1074 nm circular polarization across the observed off-limb active region. The map shown in Fig. 1C results from least squares fitting of the full Stokes profile (see Materials and Methods) at each point in the raster scanned image after first taking the median average of ten spectral rows along the slit (1.2 arcsec on sky). We recover an anti-symmetric shape of the circularly polarized Stokes V profile that is centered on the total intensity's line position (Stokes I) as shown in Fig. 2. This is strongly indicative of the Zeeman effect on the ion's fine structure energy level separation due to an external magnetic field (38). In Fig. 1C, the max amplitude of the fitted Stokes V signal in the red ling wing is shown. The red wing is selected so that its sign matches the polarity of the longitudinal field given by the weak field approximation of the Zeeman effect for an emission line (see Materials and Methods).

We measure red-wing circular polarized amplitudes between -21 to 13 $\mu B_\odot$, where the units here refer to parts per million (ppm) of the disk center spectral radiance $B_\odot$, which is ~182 photons $cm^{-2}$ $arcsec^{-2}$ $nm^{-1}$ at the observed wavelengths. Using relative units provides better intuition for the photometric dynamic range achieved by these observations. We estimate the one sigma detection limit for Stokes V to be 0.0045 $\mu B_\odot$ (*i.e.* 4.5 parts per billion of the disk radiance) based on the measurement statistics across the region (see figs. S1 and S2); however, the detection probability at any given location is intricately related to the signal amplitude and the degree of parasitic solar disk light scattered by the terrestrial atmosphere and the telescope optics into the line-of-sight (39). For these observations, the background light amplitude as a function of off-limb distance in solar radii units ($R_\odot$) can be approximated by a power law of index -0.836 and amplitude 11.6 $\mu B_\odot$, which corresponds to 80 $\mu B_\odot$ at 1.1 $R_\odot$ from solar disk center (fig. S2).

For the Zeeman effect, three key parameters dictate the amplitude of the $Fe^{+12}$ 1074 nm circularly polarized emission for an isolated parcel of coronal plasma. These are the number density of the upper level of the emitting ion, the total magnitude of the acting magnetic field, and the orientation of that field relative to the line-of-sight. It is only the component of the magnetic field parallel to the line-of-sight (also known as the longitudinal component) that induces circular polarization. The strongest Stokes V signals observed by CryoNIRSP occur near the limb at lower projected heights in the solar atmosphere where the total line emissivity is comparatively large. The total magnetic field amplitude is also presumably larger at lower heights; however, the observed signals should not be mistaken as an unbiased map of coronal longitudinal field strength, in part due to the polarized amplitude being directly influenced by the total line strength, which varies strongly across the target, unlike routinely used photospheric Zeeman-sensitive spectral lines. Only for those regions with Stokes V amplitude at least two standard deviations larger than the detection limit, we convert the V signal strength into the equivalent longitudinal magnetic field strength given by the weak field approximation in Gauss units, as shown in the first DKIST coronal "magnetogram" in Fig. 1D.

Importantly, these coronal observations do not likely probe isolated parcels of coronal gas; instead, they result from the full path integral of the polarized emissivity along a given line of sight through the structured corona. We find the influence of the magnetic topology on the circular polarization observed here manifests as three resolved lobes of common polarity signals; two negative lobes separated by a region of positive polarity (see Fig. 1C as well as the slit spectra in Fig. 2). When compared to photospheric magnetograms from the SDO Helioseismic and Magnetic Imager (HMI, 40), we find the two negative polarity lobes correlate with the corresponding negative polarity regions of the photospheric field that are in closer proximity to the limb than the leading positive flux, which suggests the signals originate from these regions. The origin of the patch of positive polarity is not as easily attributed to a photospheric counterpart; instead, we address it via modeling efforts below.

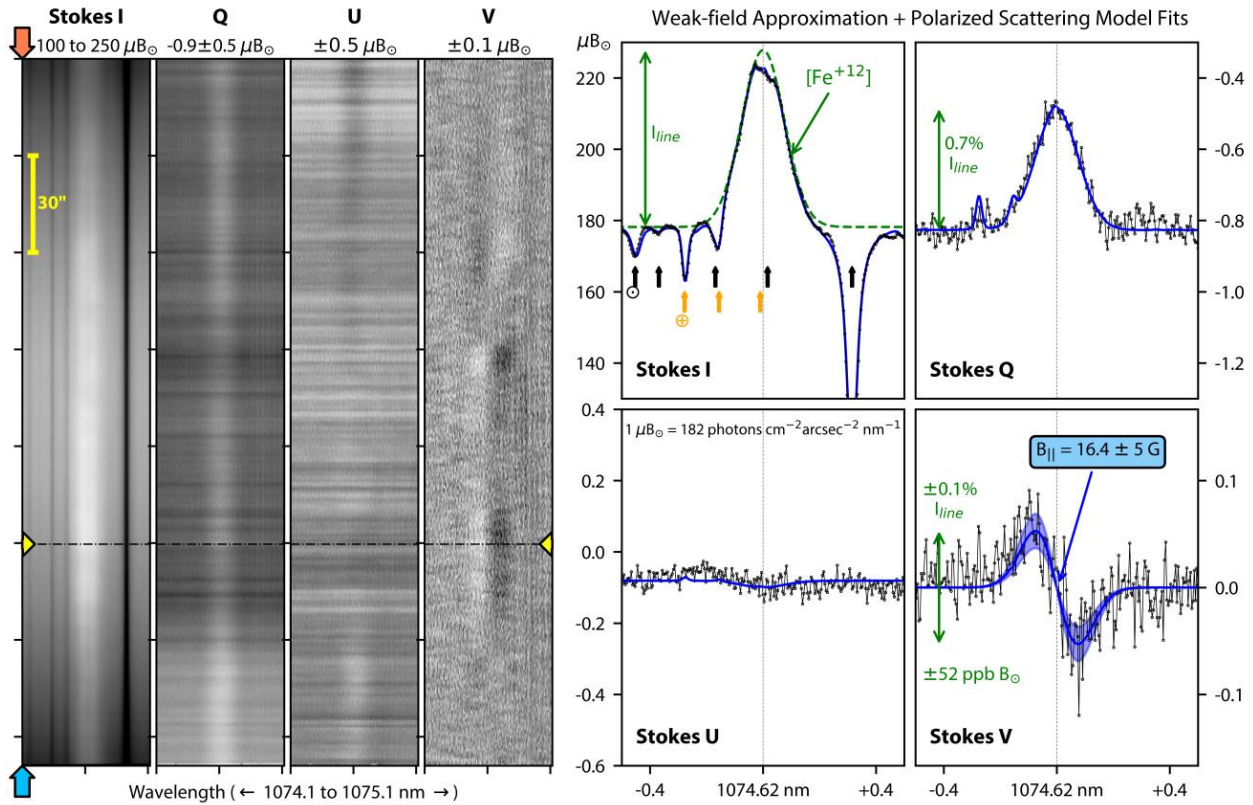

**Fig. 2. DKIST/CryoNIRSP $Fe^{+12}$ Stokes spectra and model fitted profiles.** Left hand images show the slit Stokes spectra of $Fe^{+12}$ 1074 nm obtained near the lower portion of Map A, as indicated by the blue and orange arrows in Fig. 1C. Each panel shows only a portion of the total observed spectral window. The anti-symmetric signals in Stokes V, which show a polarity flip along the slit, are the first DKIST detections of the coronal Zeeman effect. A single Stokes profile extracted along the dashed line, corresponding to the location marked with a yellow arrow in Fig. 1, is plotted in the right-hand panels together with modeled profile fits. The locations of scattered photospheric lines in Stokes I are identified with black arrows, while orange arrows indicate terrestrial absorption lines. Here, the reference direction for Stokes +Q is tangential to the slit axis and therefore, to a fair approximation, radial with respect to the Sun.

*Scattering Induced Linear Polarization*

The maps of circular polarized coronal emission that we have just presented represent a major advancement for coronal magnetometry, as only in the Zeeman induced signal is the magnetic field amplitude encoded. Meanwhile, these signals are further enhanced by the information content of the linearly polarized spectra. $Fe^{+12}$ 1074 nm linear polarization is generated through resonance scattering of the photospheric radiation field that anisotropically illuminates the coronal plasma (36, 41). The embedded magnetic field modifies the scattered radiation through the Hanle effect (42); however, the long upper-level lifetime of infrared coronal emission lines like $Fe^{+12}$ 1074 nm implies the operation of the Hanle effect saturates at negligibly small field amplitudes. Therefore, the linear polarized signal does not provide diagnostics for the magnetic field strength; instead, it probes only the orientation of the magnetic field. It is, however, generally much stronger than the Zeeman induced circular polarization (~10s of percent of the total line emission) and has subsequently been more routinely observed (43, 44). Yet, again due to the long upper level lifetimes, electron and proton collisions can efficiently depolarize the line at the higher coronal densities of the low corona (45, 46).

The high sensitivity DKIST maps of $Fe^{+12}$ 1074 nm linear polarization, which are derived from the same observations as the circular polarization maps above, are shown in Fig. 3C. Polarized amplitudes here range from 0.5 to 5% of the total line intensity (see also Stokes Q and U maps of fig. S1). Regions of enhanced $Fe^{+12}$ emission (Fig. 1B) that trace out coronal loop structures show diminished linear polarized amplitudes (<1%) in Fig. 3C consistent with their typically enhanced densities and theoretical calculations of the polarized line amplitude (32). Meanwhile, the linear polarized direction given by the line hatched overlay in Fig. 3C is visibly aligned, as expected, with the apparent coronal structure topology for the upper portions of the field-view where structures are inclined less than the Hanle critical Van Vleck angle (54.74°) with respect to the radial direction (47). For inclinations larger than the Van Vleck angle, the linear polarization direction is expected to be perpendicular to the projected magnetic field direction. Consequently, a preponderance of radial linear polarization directions is expected for off limb observations.

In addition to the emission line linear polarization, the coronal continuum spectrum is also linearly polarized. In the low corona, the continuum radiation is dominated by the K-corona created by Thomson scattering of the photospheric radiation field by free electrons (48). It is linearly polarized primarily tangential to the solar limb (49, 50). The K-corona is the dominant component of coronal emission visible to the naked eye during a total solar eclipse. The calibration accuracy required of CryoNIRSP to recover the Stokes V signals above has the additional benefit of resolving the polarized K-coronal amplitude across the regions, which is shown in Fig. 3B. We measure continuum polarization amplitudes ranging from 0.25 to 1.2 $\mu B_\odot$, which are consistent with theoretical calculations of Thomson scattering in the low corona, for which we expect the degree of polarization to be 15 to 40 percent (50). The polarized amplitude further increases within presumably denser coronal structures, which in the case includes the cusps of post flare loops that are partially visible in the AIA ultraviolet bandpass centered at 9.4 nm, shown in green within Fig. 3A. This bandpass is dominated by $Fe^{+17}$ emission broadly formed near temperatures of $10^{6.8}$ K. Within the post flare material near $\langle X, Y \rangle = \langle -125'', 1010'' \rangle$, the polarized continuum peaks at of 1.15 $\mu B_\odot$ within the orange contour, which is approximately a factor of three larger than the nearby non-active corona near $\langle X, Y \rangle = \langle -180'', 1010'' \rangle$. By estimating the loop structure depth to be equivalent to its apparent width of ~8 Mm, we can calculate, using established methods, a lower bound on its free electron density of $7 \times 10^{10}$ cm$^{-3}$ provided the emission is generated by Thomson scattering at the solar limb (51).

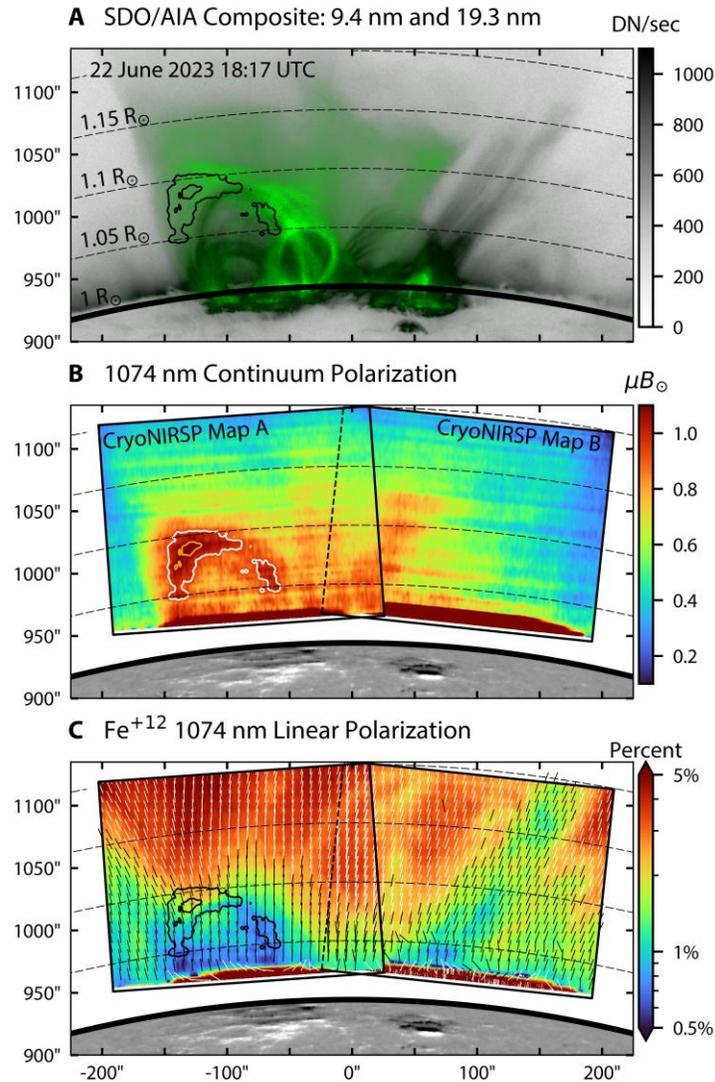

**Fig. 3. CryoNIRSP linearly polarized observations.** (**A**) A composite image blending AIA images at 19.3 nm (inverse greyscale), for which the primary contributions are $Fe^{+10}$ and $Fe^{+11}$, and 9.4 nm which is dominated by emission of $Fe^{+17}$ (green). (**B**) Linear polarized amplitude of the spectral continuum near 1074 nm, corresponding to the K-corona, in units of ppm of the disk radiance ($\mu B_\odot$). (**C**) The $Fe^{+12}$ 1074 nm emission line's linear polarization shown in units of percent of the line amplitude. Over-plotted lines denote the direction of linear polarization in this panel. Meanwhile, contour lines for continuum polarization levels of 0.95 and 1.1 $\mu B_\odot$ are shown in each panel to denote the location of enhanced polarization.

*Model-based interpretation of the Zeeman effect observations*

Our observations demonstrate the advanced new capability provided by DKIST and the CryoNIRSP instrument to map the full Stokes spectra of the $Fe^{+12}$ 1074 nm coronal emission line coronagraphically. This includes maps of the Zeeman-induced circular polarization at unprecedented temporal and spatial resolution, which holds promise for vastly improved constraints on the magnetized structure of the active solar corona. However, resolving these signals is a first step, which must be followed by concerted efforts to interpret and incorporate these diagnostics into coronal modeling efforts.

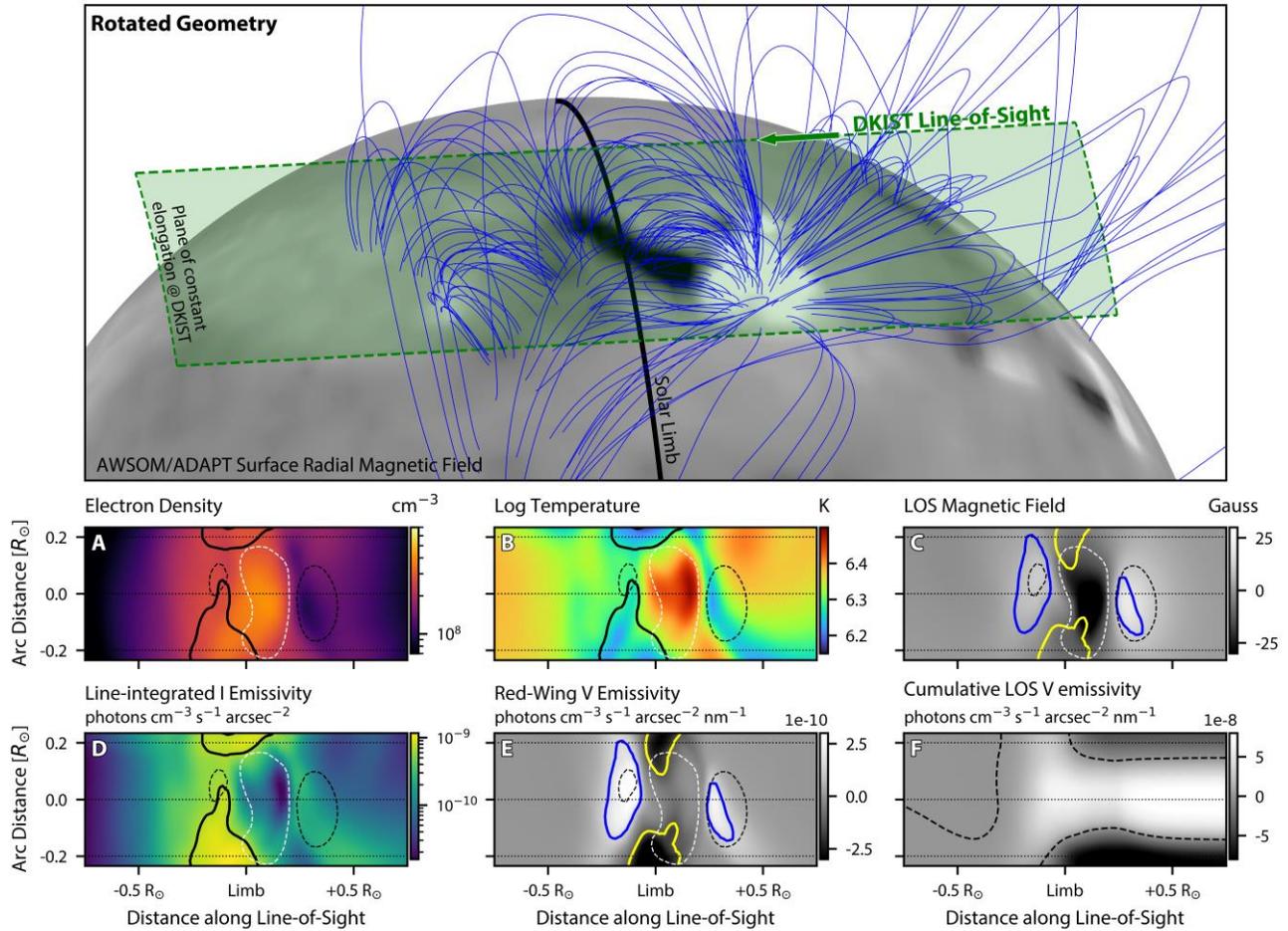

**Fig. 4. Synthetic Fe$^{+12}$ polarized emissivities through the AWSoM coronal model. (Top)** A rotated perspective of the 3D global coronal model with select magnetic field lines (blue) traced within the region observed. Solar North is towards the top of the image. The overlaid green plane is a slice of constant projected height (0.04 R$_\odot$) as observed along the line-of-sight from DKIST **(A-E)** Extracted quantities within the green plane as a function of distance from the solar limb observed from DKIST; positive distances are closer to Earth. **(F)** The cumulative Stokes V emissivity along the path increasing toward DKIST. Solar black contour lines in panels A, B, and D outline $\epsilon_I = 7 \times 10^{-10}$ photons cm$^{-3}$ s$^{-1}$ arcsec$^{-2}$. Dashed black and white contour line in A-E outline line-of-sight magnetic field strengths of ± 10 Gauss. Blue and yellow contour lines in C and E outline $\epsilon_V = \pm 1.5 \times 10^{-10}$ photons cm$^{-3}$ s$^{-1}$ arcsec$^{-2}$ nm$^{-1}$. The dashed line in panel F occurs at zero.

As a first step, Figs. 1E and 1F, together with Fig. 4, provide a model-based interpretation for the origin of the three distinct lobes of Stokes V polarization observed by CryoNIRSP for the targeted region. We compare the observations to the equivalent diagnostics forward synthesized through a global three-dimensional magnetohydrodynamic model of the corona (see Materials and Methods). For this, we adopt the Alfvén Wave Solar Atmosphere Model (AWSoM, 52, 53). A critical boundary condition driving this model is the photospheric magnetic field distribution at the specific time of interest. Due to the off limb observing geometry of the DKIST coronagraphic measurements, in addition to the lack of available complementary data from alternative vantage points on this date, we must approximate the bottom boundary using photospheric measurements of the region after it rotates onto the observable solar disk. This approach is limited by the continued evolution of the active region (fig. S3), which includes a large-scale flux emergence on 26 June 2023 that doubles the unsigned flux in the region prior to its transit of the central solar meridian. We have subsequently chosen the NSO/GONG ADAPT (54) photospheric magnetic field map from 26 June 2023, prior to the emergence, to drive the model.

The synthetic maps of Fe$^{+12}$ 1074 nm circular polarization and the corresponding synthetic magnetograms in Fig. 1E and 1F exhibit structure bearing some limited resemblance to the observations. A combination of factors is likely responsible for the substantial differences, including the crudeness of the bottom boundary used in the model and assumptions baked in by the energy equation of the model. These interesting discrepancies will need to be addressed in due course as they highlight the gap in our knowledge of the coronal structure based on these routinely used modeling approaches. That said, the synthesized data do reproduce the three-lobe character of the observed maps within the lower portion of the region; albeit the amplitudes of the circular polarization and the magnetogram are reduced in comparison to the observations. DKIST/CryoNIRSP measures absolute Stokes V amplitudes larger than 0.01 μB$_\odot$ at distances of 1.05 to 1.1 R$_\odot$, whereas the model exhibits strengths a factor of 5 to 10 lower. One explanation may be the relatively coarse ~1° resolution of the ADAPT magnetograms driving the AWSoM model, which would dilute the maximum magnetic field strengths extending from finely structured photospheric magnetic field concentrations at the base of the region (see the plage in fig. S3B).

To understand the origin of the three lobed structure of circular polarization, we investigate the formation of the polarized signals through the coronal model. The top panel of Fig. 4 depicts the 3D global coronal model from a rotated perspective. The green plane overlaid is a slice through the model along the line of sight from DKIST at the time of observation. This plane is at a constant projected height of 0.04 R$_\odot$ above the observable solar limb as viewed from DKIST, which corresponds the height of the observed slit spectrum shown in Fig. 2, though extending over the full range of position angles around the limb observed by the two CryoNIRSP maps. In Fig. 4A through 4F, select model state variables and calculated emissivities are displayed which sample the material within this plane. Black contours in Fig. 4A, 4B, and 4D surround regions of the largest line-integrated Stokes I emissivity, which help pinpoint the locations along the line-of-sight that contribute the most to the observed intensity. Meanwhile, blue and yellow contours in Fig. 4C and 4E surround regions of strong Stokes V circularly polarized emissivity in the red wing of the spectral line. Given that this quantity is signed, opposite handed circularly polarized emission along the line of sight may cancel; therefore, in panel F, the cumulative V emissivity along the path increased toward DKIST is shown as a function of distance from the solar limb.

We find the strongest contributions to the total Fe$^{+12}$ 1074 nm Stokes I emission along the line-of-sight are located directly above and just behind the solar limb as viewed from DKIST, trailing the active region core where the magnetic flux density is strongest, which has already rotated on to the solar disk viewable from Earth. Along the sampled plane, the core of the active region has the largest electron densities and longitudinal magnetic field strengths; however, the temperatures are greater than $10^{6.4}$ K, which is above the equilibrium formation temperature of Fe$^{+12}$ leading to depressed ion populations in the core and therefore low Stokes I emissivity. Similarly, the circularly polarized V emissivity is also low within the active region core. Instead, near the center of the sampled plane, which corresponds to the region of positive circular polarization observed by CryoNIRSP, the Stokes V emissivity is dominated by the positive polarity longitudinal field trailing the active region core just beyond the solar limb. In contrast, the Stokes V emission sampled by the outer portions of the sampled plane finds its dominant contributions in the oppositely directed longitudinal field just above and below the latitudes of the active region core; the latitudinal direction is along the arc of the plane sampled in the figure (see also fig. S4). In this way, we can explain the three-lobe circular polarized structure observed by CryoNIRSP (Fig. 1C); although further model refinement is needed to further substantiate these initial findings.

**Discussion**

Reconstructing the 3D distribution of coronal plasma and its embedded magnetic stresses remains essential for understanding coronal energetics. These first reported maps of the coronal Zeeman effect, made possible by the Daniel K. Inouye Solar Telescope, unveil the wealth of information that polarimetric diagnostics provide for the solar corona, particularly for its key driver: the magnetic field. The Stokes V signals that we resolve refute early preconceptions that the Zeeman Effect is too weak to reliably map the corona magnetic field. In fact, our findings suggest the longitudinal coronal field strengths are larger at greater distances above the limb ($\gtrsim$15 G at ~1.07 R$_\odot$) than routinely used global coronal models would otherwise estimate. Part of this discrepancy could be due to the lower resolution of the model used here or the crudeness of the imposed bottom boundary, which are limitations of this current study. Another question to consider, though, is whether the fundamental energy equation inherent in the model skews the thermal distribution of plasma in such a way that the regions of $Fe^{+12}$ emissivity do not adequately sample the stronger field regions of the model. This could also lead to an underestimation of the line-of-sight integrated longitudinal Zeeman effect signal. Once again, the polarized emissivities are intricately related to both the magnetic field distribution and the thermodynamic structure, which we emphasize to highlight both the complexity of these diagnostics as well as their inherent value.

Our measurements have also revealed the presence of coherent morphologies in the $Fe^{+12}$ Stokes V maps above an active region, which are in addition to complementary structures resolved in maps of $Fe^{+12}$ linear polarization. By comparing these observations with the forward synthesized observables, we have identified the probable source regions for the coherent lobes of common Stokes V polarity. It is yet to be demonstrated whether active regions routinely produce such coherent polarized morphologies when observed at the limb; however, one may already anticipate the favorable constraints such morphologies provide for the magnetic field configuration even before the associated photospheric field distribution is observable from Earth. Furthermore, we anticipate the possibility to study the evolution of these morphologies for various active regions and other structures, jointly in linear and circular polarization, as a way to diagnose the buildup of coronal free energy (55).

In summary, our work has observationally confirmed that the Zeeman effect, and more generally, large-aperture coronal polarimetry, provides powerful diagnostic tools for solar coronal research. While a number of important strategies and techniques are being advanced for coronal magnetometry, for example, the seismological use of coronal waves (44,56), radio emission (57), and unique magnetically induced EUV transitions (58), Zeeman effect observations are uniquely positioned to advance this area of research as well as help cross-validate other techniques. Even given the complexities of the optical thin formation of these observations, which may at times be overcome using tomography and single-point approximations (59–61), detections of a Stokes V signal provide a hard lower bound on the distribution of the magnetic field within the corona. Yet, more than that, high sensitivity and accuracy full Stokes polarized maps provide unique constraints on solar coronal models that are critical to the progression towards reliable physics-based modeling of space weather. The advanced large aperture coronagraphic capabilities of DKIST has opened these new possibilities.

**Materials and Methods**

*Data Description*

The Cryogenic Near-Infrared Spectropolarimeter (CryoNIRSP, 31) is a facility instrument of the US National Science Foundation's Daniel K. Inouye Solar Telescope (DKIST, 30). The CryoNIRSP data used here include two raster scans of the off-limb solar image across its 0.5 arcsec wide spectrograph slit. In each scan, the slit was oriented tangential to the solar limb, and the solar image was scanned in the direction away from the solar disk. The two maps were acquired on 22 June 2023. The first map (A) commenced at 17:42 UT and extended until 19:00 UT. The second map's (B) time ranged from 19:18 to 20:37 UT. We note that these target locations were selected based on the consensus predictions of the solar source regions connected to the Parker Solar Probe during its 16th perihelion on this date (62). During the CryoNIRSP maps, the telescope's optical axis was centered at Helioprojective coordinates of $\langle X, Y \rangle = \langle -1003'', 250'' \rangle$ and $\langle -944'', 420.5'' \rangle$ corresponding to a Helioprojective radius of 1.095 $R_\odot$. The prime field stop of the telescope rejects most of the disk light by reducing the field-of-view to a 5-arcminute diameter circle. The remaining one arcminute of the near-limb solar disk is subsequently blocked at the secondary focus with an occulting blade. A Lyot pupil stop is located between the primary and secondary foci, reducing the effective telescope collecting area by 19%.

The CryoNIRSP raster scans are comprised of a series of 46 image step positions separated by 4 arcsec. The final focal plane consists of two spectral images that exit the dual beam polarimetric analyzer. The illuminated length of the slit is 225″, sampled at 0.12″ per pixel. The effective spatial resolution is sample limited in the scanning direction (~8″); meanwhile, along the slit, it is limited to a few arcseconds by terrestrial seeing which shows some qualitative degradation over the time period of the observations. The spectral dispersion is 4.4 pm per pixel, extending over 1000 spectral pixels, with a resolving power of R ~ 48,000. At each step position, up-the-ramp exposure sequences with 10 frames (including reset and bias frames) were acquired at 16 discrete phase angles of the crystal polarimetric modulator and then repeated 7 times. The rate of modulation was 1.08 Hz with 725 msec integration per modulation state, for a total of 81.2 seconds of total integration per step. The time between adjacent slit positions was 105 seconds. The per pixel photon flux is estimated to be 560 photons per second per $\mu B_\odot$. Thus, for a 180 $\mu B_\odot$ background scattered light level, as in Fig. 2, we would expect a root-mean-squared (rms) noise level of ~19 ppb (for photon errors) for the full processed polarimetric data assuming dual beam combination and averaging of 10 pixels along the slit. For the coronal line spanning approximately 50 spectral pixels, the expected rms error reduces to ~4 ppb, which is consistent with the noise floor found in our fitting results (see fig. S2).

*Data Processing*

Achieving a high signal to noise ratio, in addition to high polarimetric accuracy, for this data requires careful mitigation of specific observational artifacts. Standard processing includes detector, geometric and polarimetric calibration steps (63, 64); however, residual crosstalk between the Stokes parameters can still be present. Interference fringes also impact the data. A Fourier amplitude analysis of flat-field calibration data identifies 8 dominant fringe periods (6.3, 8.7, 11.8, 13.4, 19, 22.2, 24.8, 26.4, 34.4 and 44 spectral pixels) with amplitudes between 0.02% and 0.5%. These are attributed to transmissive optics in the optical path, particularly the crystal modulator, which has since been upgraded to further improve the instrument performance. For these data, we devise an iterative scheme that filters the fringes in frequency space using a multiple bandstop Butterworth filter while also determining the Stokes I to Q, U, and V residual crosstalk by minimizing the presence of scattered photospheric lines in the polarized spectra (50). A demonstration of these steps, which are done prior to dual beam combination, is shown in fig. S5 for the Stokes V slit spectra shown in Fig. 2. We also implement a correction for residual linear to circular crosstalk, as done by previous authors (29), which is determined during Stokes profile fitting discussed below.

*Coronal Stokes Spectral Model Fitting*

The theoretical basis for polarized coronal line formation within forbidden emission lines is well established (36, 65). The joint influence of resonance scattering, the saturated Hanle effect, collisional depolarization, and the Zeeman effect can be well described using a multi-level model of the ion within the atomic density matrix formalism. Under the assumption of a Maxwellian distribution of ion velocities, the line profile for Stokes I and the linear polarized (Q and U) emissivities take on a Gaussian shape. Meanwhile, the circular polarized profile is given by the derivative of Stokes I under the weak field approximation (36). Here we ignore the influence of atomic alignment on Stokes V, which becomes negligibly small in denser regions of the corona. To perform least squares fitting of the observed Stokes spectra, we start with a model of the Stokes parameters as follows:

$$I(\lambda) = b I_{disk}(\lambda) + G_E(\lambda; a, \lambda_0, \sigma) \qquad (1)$$

$$Q(\lambda) = cos(2\alpha)\, p\, G_E + Q_c \qquad (2)$$

$$U(\lambda) = -sin(2\alpha)\, p\, G_E + U_c \qquad (3)$$

$$V(\lambda) = -\left(\frac{\lambda_0^2 e_0}{4\pi m_e c^2}\right) \bar{g} B_{||} \left(\frac{\partial I}{\partial \lambda}\right) + X_{P2V}(p\, G_E) \qquad (4)$$

where $b$ is the fraction of photospheric solar light $I_{disk}$ scattered into the light of sight; $G_E$ is the Gaussian coronal line profile with amplitude $a$, line center $\lambda_0$, and line width $\sigma$; $p$ is the linear polarized fraction; $\alpha$ is the angle of linear polarization; $Q_c$ and $U_c$ are the continuum linear polarization due to Thomson scattering of the K-corona. $e_0$ is the elementary charge and $m_e$ is the electron mass. $\bar{g}$ is the effective Landé factor. $B_{||}$ is the longitudinal component of the magnetic field. Finally, $X_{P2V}$ is added as a correction factor for residual crosstalk from linear to circular polarization.

When fitting the above model with least squares optimization, we further include the effects of terrestrial atmospheric extinction and convolution with the spectrograph line spread function ($L$) on each parameter; for example, $I \Rightarrow L(\sigma_i) * I\, e^{-\tau_\oplus}$ where $L$ is approximated as a Gaussian function and $\tau_\oplus$ is the optical depth of terrestrial absorbers. $I_{disk}$ is approximated by the Solar Pseudo-Transmittance Spectrum provided by TCCON network (66). $\tau_\oplus$ is calculated using the HITRAN molecular database for water absorption and the Python for Computational Atmospheric Spectroscopy code (Py4CAtS, 67, 68). We scale the water absorbance linearly during the fit using an additional fit parameter to account for airmass variations. We demonstrate the various components of the least squares fitting approach in fig. S6. The derived parameters for the CryoNIRSP raster scans A and B are shown in Fig. 1 and fig. S1.

*The Coronal Global Model and Forward Line Synthesis*

Our initial interpretation of the observed polarized signals is based on forward-synthesized polarized intensities emergent from global magnetohydrodynamic models of the solar corona. Preliminary investigations with routinely produced model runs from the Predictive Sciences Inc. Magnetohydrodynamics Around a Sphere model (69), as well as the Alfvén Wave Solar atmosphere Model (AWSoM, 52, 53) both yielded some favorable correspondence with the data despite using traditional synoptic magnetograms as the bottom boundary. Using the AWSOM model, we perform further refinement of the bottom boundary conditions for the analysis reported herein (70). AWSoM is a three-dimensional magnetohydrodynamic model of the solar corona that is driven at the lower boundary by the observed photospheric magnetic field distribution and is self-consistently heated via low-frequency Alfvén wave turbulence. We refer the reader to the primary AWSoM reference papers for more information (52, 53). We adopt the NSO/GONG ADAPT magnetogram from 26 June 2023 for the bottom boundary, which assimilates data of the targeted active region 3 days after it was observed off-limb by CryoNIRSP. The MHD equations are solved on a spherical block adaptive grid (70) extending from 1 to 24 $R_\odot$. A total of 4,624,896 grid cells are used in this case. The highest angular grid resolution is 1.4° near the bottom boundary ($\lesssim 1.1$ $R_\odot$), while the radial sampling occurs on a stretched grid. The mean radial cell size increases from $10^{-4}$ $R_\odot$ at the bottom boundary, to 0.02 $R_\odot$ at 1.15 $R_\odot$, to 0.86 $R_\odot$ at 20 $R_\odot$.

Using the 3D state variables from AWSoM, we calculate the polarized Stokes emissivities for the $Fe^{+12}$ 1074 nm line throughout the model domain using the Python package for Coronal Emission Line Polarization (32, 72). The logarithmic iron abundance relative to $10^{12}$ hydrogen atoms is set to 8.1 dex (73). The atomic data is sourced from the CHIANTI database V10.1 (74), but we truncate the number of atomic levels to the 200 lowest energy states for numerical expediency. The reduced number of levels provides sufficient accuracy for our purposes (32). We also do not include macroscopic Doppler motions in these calculations. The full Stokes emissivities $\epsilon_{I,Q,U,V}$ are calculated; however, only $\epsilon_I$ and $\epsilon_V$ are discussed here. $\epsilon_V$ includes the influences of atomic alignment that are not represented by the weak field approximation of the Zeeman effect. In Fig. 1, however, we fit the synthesized Stokes V spectra profiles with the weak field approximation to give a Gauss equivalent to the longitudinal circularly polarized amplitude in the same manner applied to the observations.

**Acknowledgments**

The research reported herein is based in part on data collected with the Daniel K. Inouye Solar Telescope (DKIST), a facility of the National Science Foundation. DKIST is operated by the National Solar Observatory under a cooperative agreement with the Association of Universities for Research in Astronomy, Inc. DKIST is located on land of spiritual and cultural significance to Native Hawaiian people. The use of this important site to further scientific knowledge is done with appreciation and respect. This work utilizes data produced collaboratively between Air Force Research Laboratory (AFRL) and the National Solar Observatory (NSO). The ADAPT model development is supported by AFRL. The input data utilized by ADAPT is obtained by NSO/NISP (NSO Integrated Synoptic Program). AWSoM simulations were performed on Pleiades of the NASA Advanced Supercomputing (NAS) Division. The SDO data are provided courtesy of NASA/SDO and the AIA and HMI science teams.

**Funding:**
The Daniel K. Inouye Solar Telescope (DKIST) is funded by the National Science Foundation (Award #1400450).

**Author contributions:**
Conceptualization: TS, GP, AF, JK
Investigation: TS, JK, AF, IS, RW, TR, AT, FW, DH
Methodology: TS, JS
Formal Analysis: TS
Writing—original draft: TS, GP, AP
Writing—review & editing: TS, GP, JK, AF, TR, AT, FW, IS, RW, DH, AP, JS

**Competing interests:**
Authors declare that they have no competing interests.

**Data and materials availability:**
All data needed to evaluate the conclusions in the paper are present in the paper and/or the Supplementary Materials.

**Supplementary Materials**

Figs. S1 to S6.

# Supplementary Materials for

**Mapping the Sun's coronal magnetic field using the Zeeman effect**

Thomas Schad *et al.*

*Corresponding author. Email: tschad@nso.edu

**This PDF file includes:**

Figs. S1 to S6

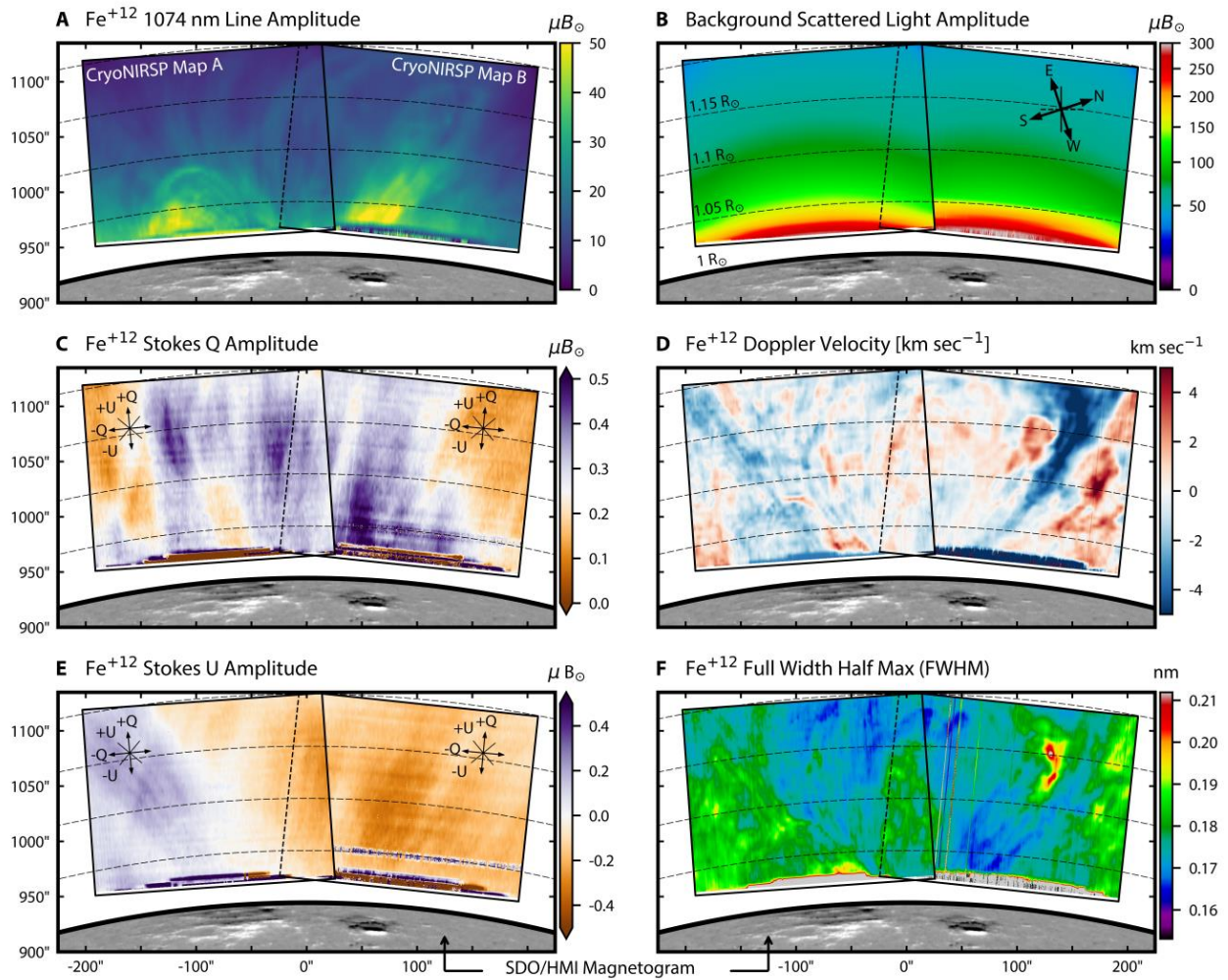

**Fig. S1. Maps of parameters derived from full Stokes profile fits of Fe$^{+12}$ 1074 nm.** (**A**) Peak Stokes I Gaussian line amplitude. (**B**) The total amplitude of the background photospheric scattered light in the spectral continuum across the field of view. Solar cardinal directions are as shown. (**C**) and (**E**) The signed maximum Stokes Q and U line amplitudes in a Stokes reference frame where +Q is tangential to the slit axis for each map, *i.e.* approximately radial. (**D**) Doppler velocity of the Fe$^{+12}$ 1074 nm relative to the median observed line center position of 1074.62 nm. (**F**) Full width half maximum line width of the coronal emission line.

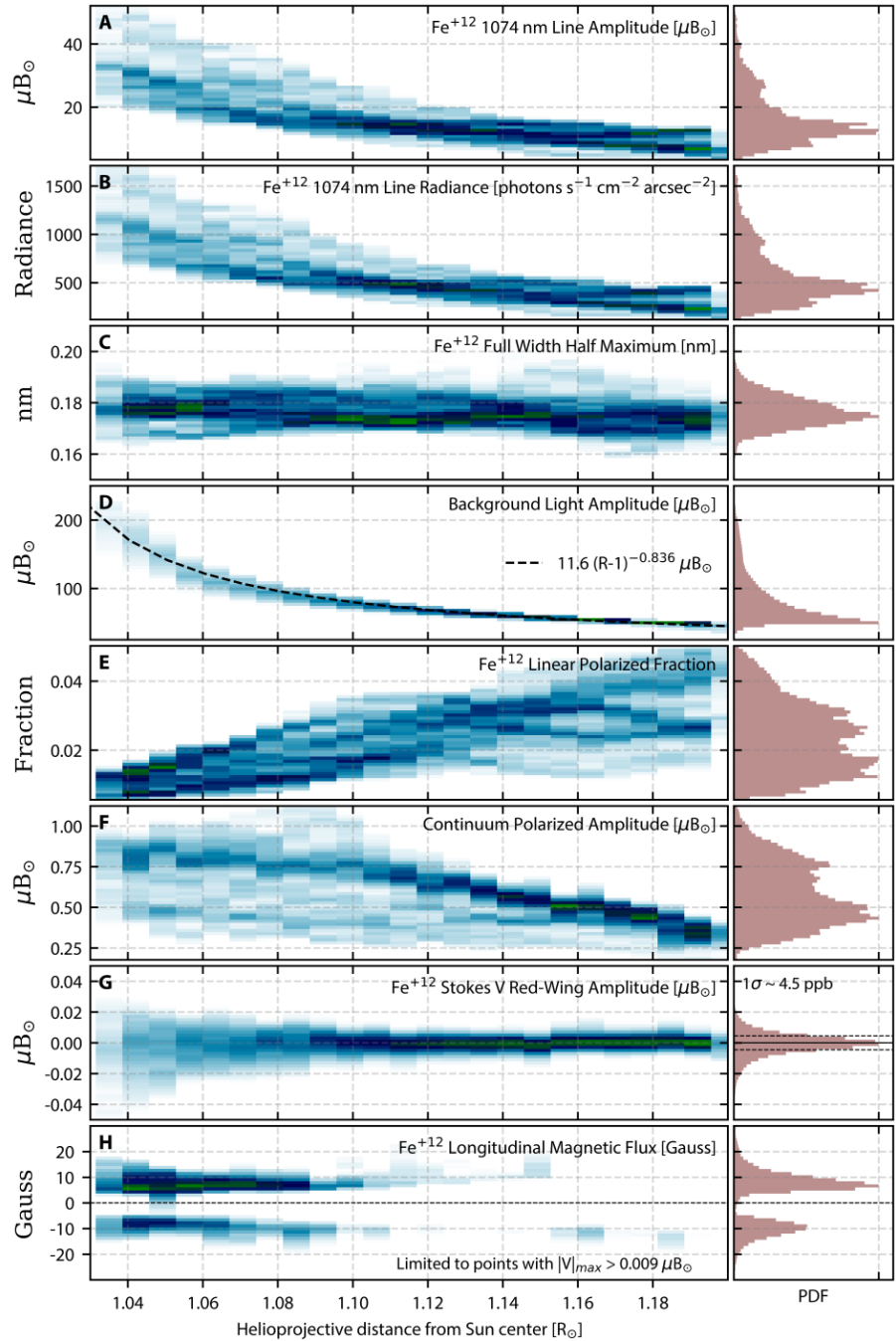

**Fig. S2. Histograms of derived $Fe^{+12}$ 1074 nm line parameters.** The left-hand columns provide 2D histograms of each variable with respect to the Helioprojective distance from Sun center. The histograms include all $10^6$ points of the two CryoNIRSP raster maps except for panel H, which restricts the points to those with maximum unsigned Stokes V polarization larger than two times the estimated $1\sigma$ noise level. The right-hand panels show the corresponding 1D histogram normalized to its peak value.

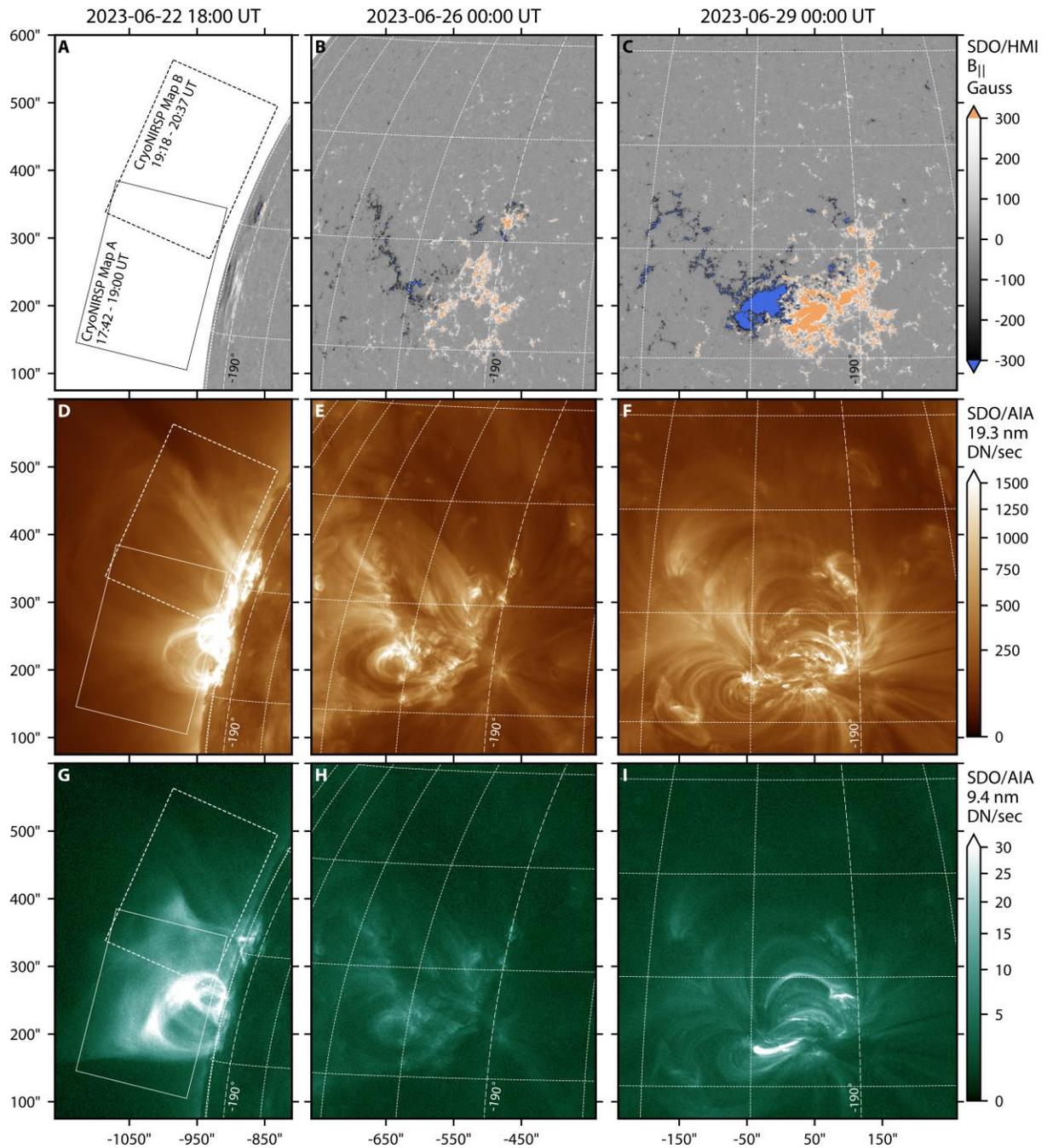

**Fig. S3. Evolutionary context of the targeted active region.** **(A-C)** SDO/HMI longitudinal magnetograms showing the photospheric magnetic field. **(D-F)** SDO/AIA 19.3 nm images **(G-I)** SDO/AIA 9.4 nm images. The left most column panels correspond to the time of the DKIST/CryoNIRSP observations with the two raster positions indicated in panel A. The middle column is 3 days 4 hours later and corresponds to the time for which the ADAPT magnetogram is selected for modeling the corona with AWSoM. The right most images show the region near the central meridian after the large flux emergence event that occurred on 26 June 2023. Heliographic Carrington latitude and longitude lines are overplotted in each panel.

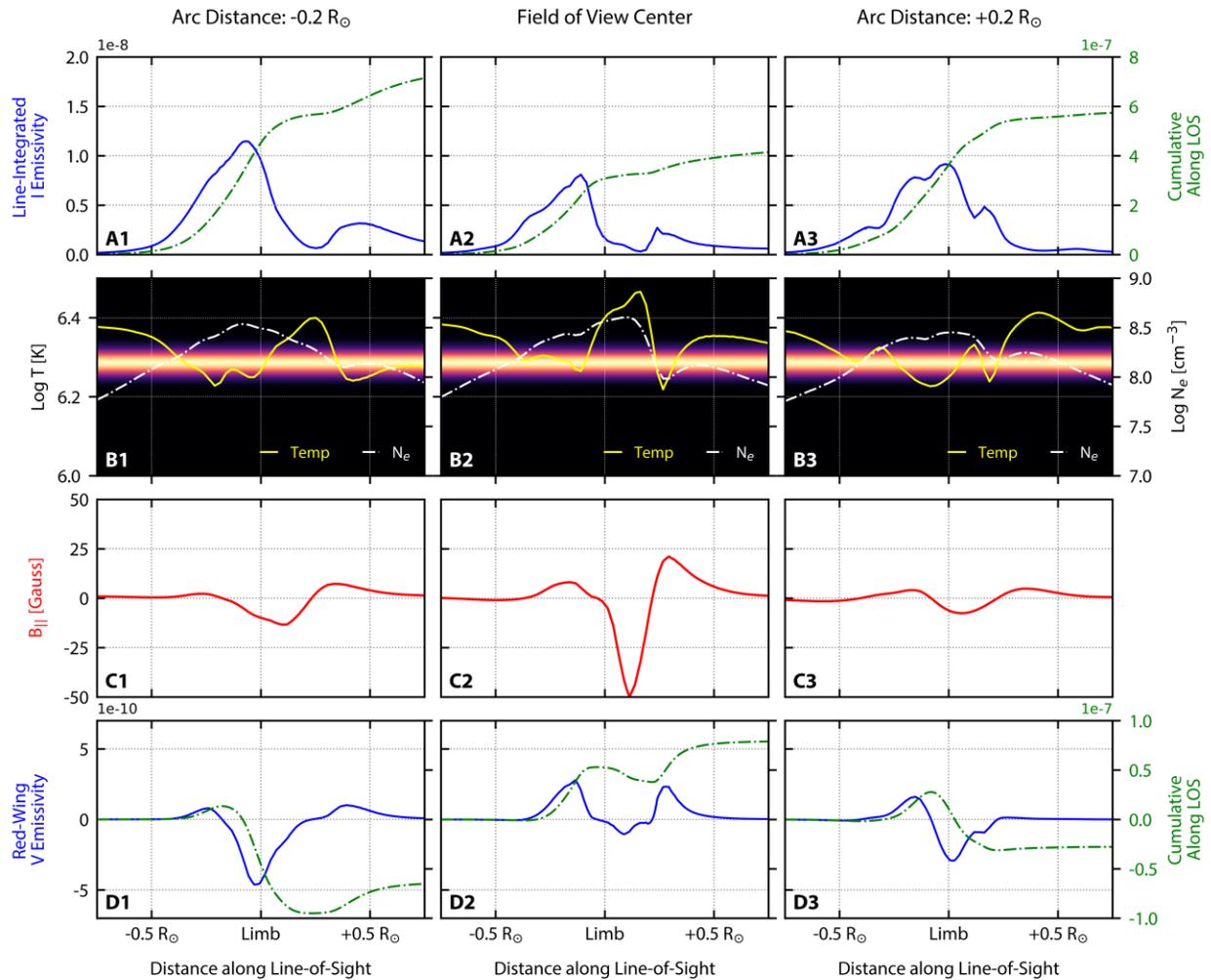

**Fig. S4. Modeled state variables and polarized emissivities along three observed lines of sights.** The left, middle and right columns show profiles of quantities modeled along the horizontal dotted lines in the lower panels of Fig. 4. These three examples correlate with the three coherent lobed structures of Stokes V observed by CryoNIRSP, with negative, positive, and negative polarities, respectively. I and V emissivities are shown with units of photons cm$^{-3}$ s$^{-1}$ arcsec$^{-2}$ and photons cm$^{-3}$ s$^{-1}$ arcsec$^{-2}$ nm$^{-1}$. Cumulative profiles in panels A and D are simple sums over the extracted profile from left to right in the panels, corresponding to increasing proximity to Earth. The yellow line in the B panels show the electron temperature profile and the background image uses an increasing brightness color map to represent the temperature-dependent Fe$^{+12}$ intensity contribution function, i.e. the temperature range over which Fe$^{+12}$ emits. Meanwhile, the C panels show the longitudinal magnetic field amplitude along the line of sight. It can be seen from these profiles how the thermal structure and magnetic field properties together dictate the observed Stokes V polarity and its amplitude.

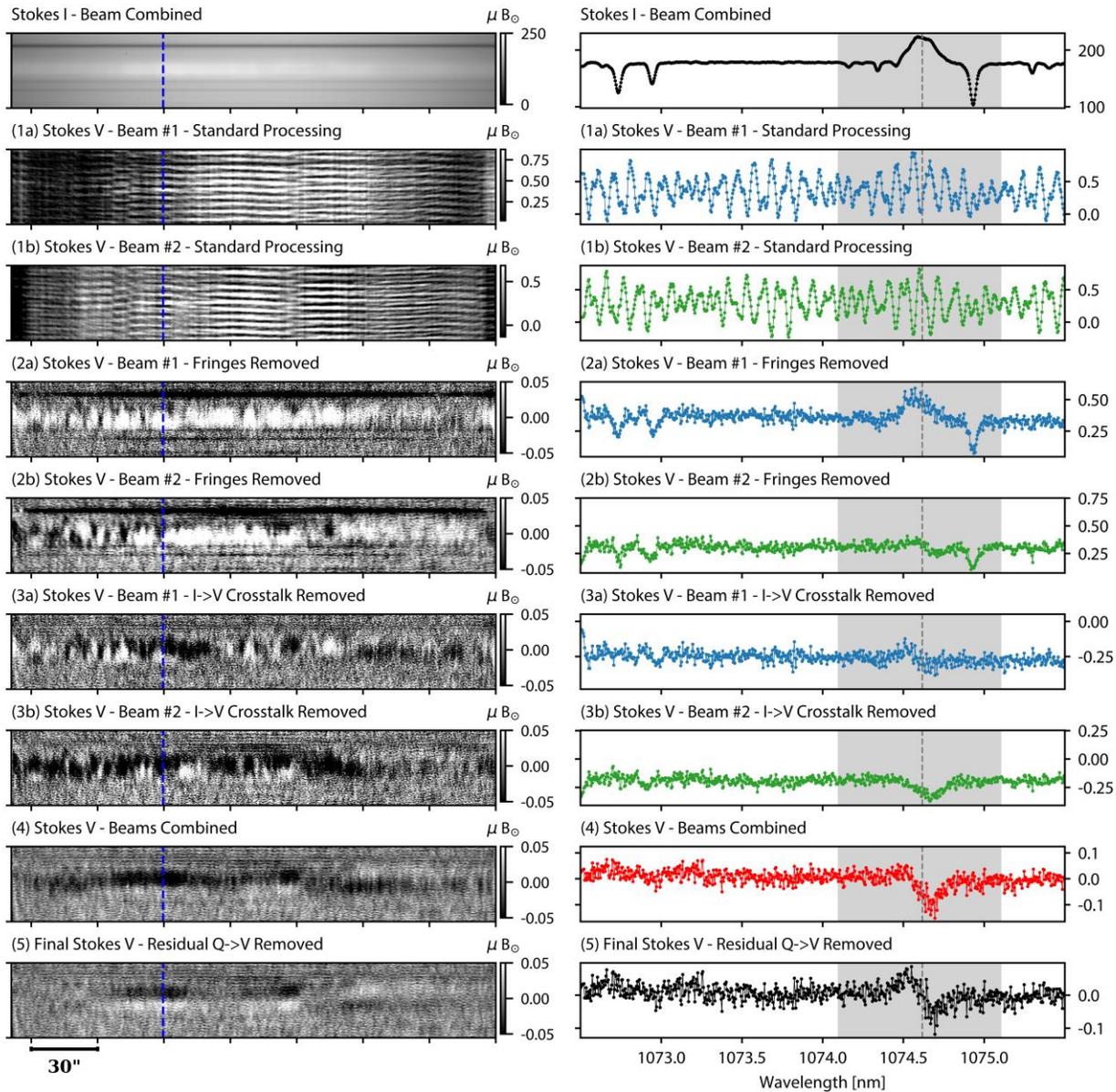

**Fig. S5. Data processing steps including interference fringe and crosstalk mitigation.** Image spectra (left) and line profiles (right) extracted from Map A at the same locations as in Fig. 2. The image spectra vertical and horizontal axes correspond to wavelength and spatial distance in arcseconds along the slit, respectively. Top to bottom panels display the intermediate processing steps starting with the standard calibrated individual polarimetric beams labeled (1a) and (1b) and finishing with the final Stokes V spectra (5). Stokes Q and U processing includes the same steps up to and including dual beam combination (4). For display purposes, panels 2a through 5 are shown with the median value of each spectral profile subtracted. In addition, color scales ranges for left hand panels may be saturated to increase contrast. The wavelength range shown in the left-hand panels is indicated at right by the grey background.

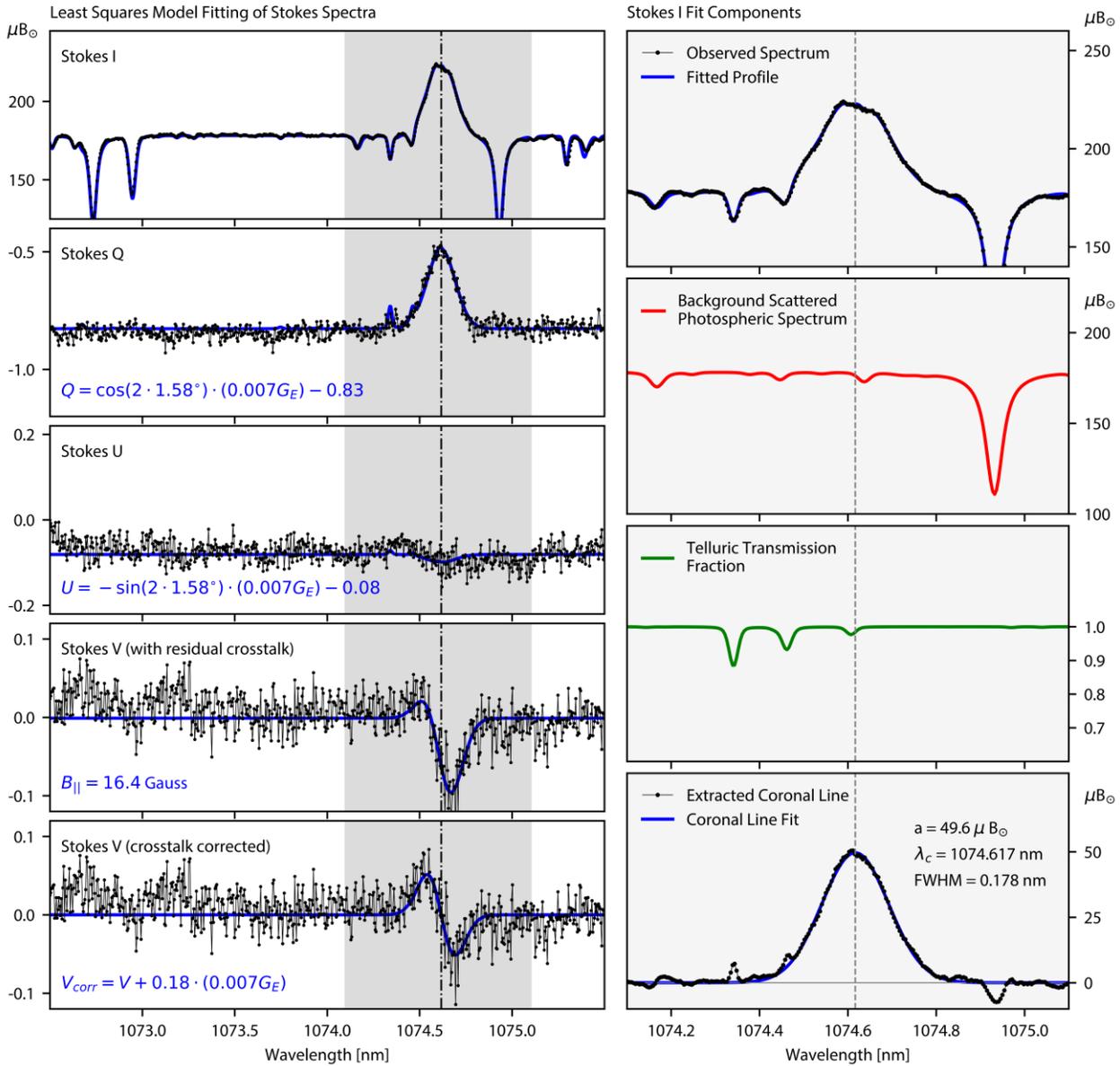

**Fig. S6. Full Stokes spectra least squares model fitting.** Left panels show the extended spectral range model fits for Stokes I, Q, U, V for the same profiles as shown in Fig. 2. The Stokes V profile before and after the correction for crosstalk between linear and circular polarization is shown. A zoomed in view of the sub-region of Stokes I is shown in the right-hand panels to display the individual components contributing to the line fit.